\documentstyle[12pt,axodraw]{article} 
\def\pri{^{\, \prime}}

\def\prd#1{{\em Phys.~Rev.}~{\bf D#1}\ }

\def\prl#1{{\em Phys.~Rev.~Lett.}~{\bf #1}\ } 
 
\def\np#1{{\em Nucl.~Phys.}~{\bf B#1}\ }
\def\deg{\ifmmode{^{\circ}}\else ${^{\circ}}$\fi} 
\def\ni#1{\noindent$(#1)\quad$}

\def\lsim{\,\raisebox{-0.13cm}{$\stackrel{\textstyle<}{\textstyle\sim}$}\,} 
\def\bi{\begin{itemize}}
\def\ei{\end{itemize}} 
\def\ed{\end{document}}
\def\be{\begin{equation}} 
\def\ee{\end{equation}}
\def\bea{\begin{eqnarray}}
\def\eea{\end{eqnarray}}
\def\req#1{(\ref{eq:#1})}
\def\eq#1{Eq.~(\ref{eq:#1})}
\def\labeq#1{\label{eq:#1}} 
 
\def\tfrac#1#2{{\textstyle\frac{#1}{#2}}}

\def\kev{\ \mbox{KeV}}

\def\eb{\end{thebibliography}}

\def\nn{\nonumber}

\def\labeq#1{\label{eq:#1}}
\def\req#1{(\ref{eq:#1})}
\def\eq#1{Eq.~(\ref{eq:#1})}
\def\tr{\ifmmode{\mbox{Tr}}\else Tr\fi} 
\def\bb{\bibitem}

\def\co#1{${\cal O}(#1)$}
\def\ri{\vec r_i}
\def\rj{\vec r_j}
\def\vpi{\vec p_i}
\def\ai{\vec A(\ri)}

\def\sij{\sum_{i\ne j=1}^{N}}
\def\sll{\sum_{\lambda,\lambda\pri}}
\def\vk{\vec k}
\def\vkp{\vec k\pri}
\def\ok{\omega_k}
\def\okp{\omega_{k\pri}}
\def\op{\omega_{pl}}
\def\ak{a_{\vk \lambda}}
\def\akd{a_{\vk \lambda}^{\dagger}}
\def\ek{\vec\epsilon_{\vk\lambda}}
\def\ekp{\vec\epsilon_{\vkp\lambda\pri}}
\def\ekek{\left|\ek\cdot\ekp\right|^2}
\def\dkk{\int\frac{d^3\vk}{(2\pi)^3 2\ok}\int\frac{d^3\vkp}{(2\pi)^3 2\okp}}
\def\dkkt{\int\frac{d^3\vk}{(2\pi)^3 2{\ok}^2}\int\frac{d^3\vkp}{(2\pi)^3
2{\okp}^2}}
\def\vrij{\vec r_{ij}}
\def\rij{r_{ij}}
\def\iz{\int_0^\infty}
\def\pol{\left(1+(\hat{k}\cdot\hat{k\pri})^2\right)}
\begin{document}
\begin{titlepage} 
\begin{flushright}  {\sl NUB-3172/98-Th}\\  
{\sl Feb 1998}\\  
hep-ph/9802417
\end{flushright} 
\vskip 0.5in
 
\begin{center}  
{\Large\bf Evidence for a Non-Zero Mass in the Dispersion 
Relation of Transverse Photons in the Stellar Core}\\
[.5in]   
{Haim Goldberg}\\[.1in]  
{\sl Department of Physics}\\   
{\sl Northeastern University}\\ 
{\sl Boston, MA 02115}  
\end{center} 
\vskip 0.4in

\begin{abstract}
The electromagnetic two-body interaction energy is modified in the thermal
environment of the solar core. The modifications are shown to imply a
gross violation of standard solar physics unless the dispersion 
relation $\omega_t(k)$ for the transverse
modes of the photon is altered to include a transverse mass.  A non-zero
transverse mass $m_t^2\simeq e^2 n_e/m_e$ is, of course, theoretically 
predicted to exist; the 
discussion in this paper implies that such a mass is {\em required}  
in order to maintain the success of the standard solar model. In
that sense one may claim observational evidence 
for a non-trivial dispersion relation for  transverse 
photon frequencies in the solar core.
\end{abstract} 
\end{titlepage} 
\setcounter{page}{2} 
\section{Introduction}
Electrodynamics (or gluodynamics) in an environment with non-zero 
temperature or
chemical potential is a highly explored topic in modern physics, with
references too numerous to list.   
Historically, one
of the first consequences of such considerations \cite{Tonks} 
was a prediction of 
long-wavelength  oscillations in a neutral plasma with a frequency 
\be
\op=\sqrt{\frac{e^2n_e}{m_e}}\ \ ,
\labeq{op}
\ee
where $n_e$ is the electron density. This is an entirely classical effect --
essentially a density oscillation of the electron gas which, through
coupling via 
Gauss's Law, implies a threshold frequency $\op$ in the propagation of the
longitudinal electric field component $\stackrel{\rightarrow}{\nabla}
\cdot\stackrel{\rightarrow}{E}.$ 
In passing from the 
classical to the quantum realm, the plasmon mass $\hbar\op$
becomes a zero-momentum energy threshold in a quantized spectrum for 
collective excitations in the plasma.
Experiments verifying the quantum nature of the longitudinal plasmon, 
in which 
electrons scattered from thin metallic films show energy losses in 
multiples of
$\hbar\op,$ were performed many years ago \cite{Powell}. 

As a result of interactions with  electrons in the plasma, 
the dispersion relation
$\omega_t(k)\equiv\ok$ of the transverse components of the photon 
also is altered. For small wave vector $k$, it takes
the form \cite{Tsytovich,Ruderman,Braaten}
\be
\ok^2\simeq k^2+\op^2\ \ .
\labeq{okp}
\ee
An  extension to the quantum realm of astrophysical significance 
was suggested by Adams {\em et al}\ \cite{Ruderman}: with
$\hbar\op$  interpreted as a  mass for the photon quasiparticle,  
its decay to a neutrino-antineutrino  pair becomes kinematically 
possible, and
could arguably constitute an important mechanism for energy loss 
in hot dense
stellar interiors prior to supernova. It will be some time 
before this idea can
be reliably tested.

The  transformation of a quantum-relativistic 
elementary field particle to a
quasiparticle through interaction with the environment is a concept 
central to many
systems of contemporary interest, such as the quark-gluon  and the 
electroweak plasmas, and the theoretical structures required to 
describe these
systems have been  extensively developed \cite{Dolan}. 
It is therefore interesting that there seems as yet to exist 
little or no {\em
observational} support for these ideas. It is the purpose of 
this work to present in a quantitative manner some evidence, albeit indirect, 
for the  necessity to use  a transverse
electromagnetic field with the modified dispersion relation
\req{okp} in order to describe quantum processes (such as scattering) 
in the stellar core.

In order to demonstrate that propagators for the quanta 
of the interacting electromagnetic field require the 
dispersion relation \req{okp}, one necessarily must 
focus on  properties in the environment which involve Feynman graphs (or the
equivalent) as contrasted with  those solely
involving frequencies of classical wave propagation. In this work I examine  
radiative corrections to the electromagnetic energy density in the 
core of the sun. 
I will  show that in the thermal environment there are quantum 
contributions of  \co{e^4} to this energy
which, for massless transverse photons, are 
phenomenologically unacceptable. 
If, on the other hand,  the
transverse photon indeed behaves like a  quasiparticle with 
mass given by $\hbar\op,$
these corrections turn out to be negligible, the energy per particle
amounting to no more than $\sim 10^{-5}\ T_c,$ where $T_c\simeq 1.3\
{\rm KeV}$ is a typical solar core temperature. It is this argument 
that I take
as (perhaps the first) observational 
evidence for the existence of transverse quanta with the 
dispersion relation \req{okp}.

\section{The Two-Particle Electromagnetic Force in a Thermal Background}
Many years ago, Tryon \cite{Tryon} considered the scattering of a charged 
particle from an external 
potential in a thermal environment, and calculated \co{e^2}\  
corrections to the cross
section due to the presence of background photons. The calculation served as
the basis for a speculation in the concluding remarks of that paper 
that (uncalculated)
infrared-finite terms might give rise to interesting phenomena in stellar
interiors. In what follows I will calculate a set of thermal  
corrections to the
{\em static} infrared-finite two-particle Coulomb energy, and assess their 
effect on stellar
physics. These corrections will {\em not} cancel because of overall charge
neutrality.

In the non-relativistic limit, the Hamiltonian describing the 
interaction of a
set of $N$ charged particles  with coordinate  vectors $\vec r_i,$ 
momentum operators 
$\vec p_i,$
charges $e_i,$ and masses $m_i$ with the quantized electromagnetic 
field $\vec
A(\vec r)$ (in Coulomb gauge) is
\be
H_{em}=\sum_{i=1}^{N}\frac{(\vpi-e\ai)^2}{2m_i}\ \ + \frac{1}{2}\sum_{i\ne j
=1}^{N}\frac{e_i e_j}{|\ri-\rj|}\ \ ,
\labeq{ham}
\ee
where the Schr\"odinger field  operator $\ai$ has the familiar 
wave number expansion
\be
\ai=\sum_{\vk} \sqrt{\frac{\hbar}{2\omega_kV}}\ \ek\ 
\left(\ak e^{i\vk\cdot\ri} + 
\akd e^{-i\vk\cdot\ri}\right)\ \ ,
\labeq{ai}
\ee
in a real basis for the polarization vector $\ek$.
In \eq{ham} and in what follows, I will omit factors of $c$ 
but retain   $\hbar$'s 
so that  the
quantum properties of various contributions are apparent.

To \co{e^2} and in the
non-relativistic limit, the two-body force between pairs of particles \{$i,j$\} 
will consist of the
(Debye-screened) static Coulomb force and, to \linebreak
\co{\vec v_i\cdot\vec v_j/c^2}, the 
magnetic Biot-Savart force (which averages to zero).
In \co{e^4}, aside from simple rescatterings, there
are processes, examples of which are  shown in the time-ordered graphs of 
Fig. 1, which make new 
contributions to the
two-body energy. 

The thermal environment acts both directly, by
allowing photons to be absorbed and re-emitted (as in Figs. 1$(b),(c)$ and
$(f))$, 
and
indirectly, by enhancing the emission and re-absorption of photons in the
intermediate state (in all the graphs). In addition, in Coulomb gauge, the
diagrams may be classified as containing either  one or more {\em kinetic}
vertices from the $\vec p\cdot \vec A$ terms of \eq{ham}, or  only
{\em static} contributions from the ${\vec A}\ ^2$ term.
Examples of diagrams with kinetic contributions are 
shown in Figs. 1$(d$-$f).$ As discussed in the concluding remarks, the
results of this paper are fully illustrated by considering only the 
static
contributions: thus, the calculation is limited to an evaluation of Figs.
1$(a),(b),$ and $(c).$  It is also clear from the mass in the 
denominator of the
static term that only the interaction among electrons need be considered.
\begin{center}
\SetScale{0.7}
\begin{picture}(540,300)(0,0)
\SetOffset(70,200)
\GBox(0,0)(1,132){0}
\GBox(55,0)(56,132){0}
\Vertex(0,38.5){1.5}
\Vertex(55,93.5){1.5}
\PhotonArc(55,38.5)(55,90,180){1.5}{12}
\PhotonArc(0,93.5)(55,270,360){1.5}{12}
\GBox(180,0)(181,132){0}
\GBox(235,0)(236,132){0}
\Vertex(180,93.5){1.5}
\Vertex(235,38.5){1.5}
\Photon(150,25)(180,93.5){1.5}{12}
\Photon(180,93.5)(235,38.5){1.5}{12}
\Photon(235,38.5)(265,107){1.5}{12}
\GBox(360,0)(361,132){0}
\GBox(415,0)(416,132){0}
\Vertex(360,42){1.5}
\Vertex(415,90){1.5}
\Photon(340,5)(360,42){1.5}{6}
\Photon(360,42)(415,90){1.5}{12}
\Photon(415,90)(435,127){1.5}{6}
\Text(23,-15)[]{$\small a$}
\Text(150,-15)[]{$\small b$}
\Text(277,-15)[]{$\small c$}
\SetOffset(70,70)
\GBox(0,0)(1,132){0}
\GBox(55,0)(56,132){0}
\Vertex(0,38.5){1.5}
\Vertex(55,93.5){1.5}
\Vertex(0,56.5){1.5}
\Vertex(55,75.5){1.5}
\Photon(0,38.5)(55,93.5){1.5}{12}
\Photon(0,56.5)(55,75.5){1.5}{8}
\GBox(180,0)(181,132){0}
\GBox(235,0)(236,132){0}
\Vertex(180,38){1.5}
\Vertex(235,66){1.5}
\Vertex(180,90){1.5}
\Photon(180,38)(235,66){1.5}{10}
\Photon(180,90)(235,66){1.5}{10}
\GBox(360,0)(361,132){0}
\GBox(415,0)(416,132){0}
\Vertex(360,38.5){1.5}
\Vertex(360,90){1.5}
\Vertex(415,70){1.5}
\Vertex(415,50){1.5}
\Photon(330,5)(360,90){1.5}{12}
\Photon(360,35)(415,75){1.5}{9}
\Photon(415,50)(450,127){1.5}{12}
\Text(23,-15)[]{$\small d$}
\Text(150,-15)[]{$e$}
\Text(277,-15)[]{$f$}
\end{picture}\end{center}\vspace*{-1cm}

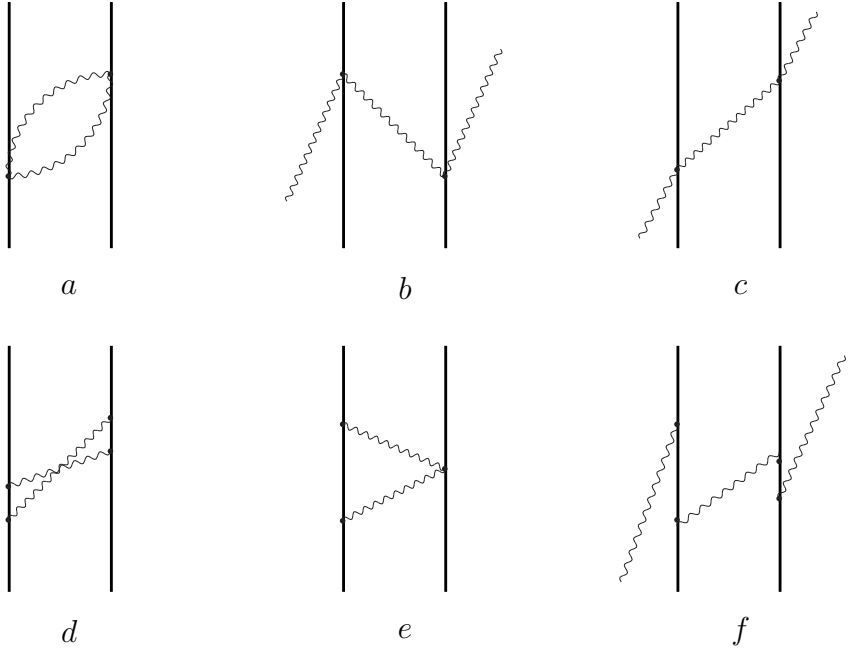
\hspace*{0.3in}\parbox{5in}{\noindent \small Figure 1: Various 
time-ordered graphs contributing to the interaction
energy in a thermal environment. The graphs $(a),(b),(c)$ provide the static
contributions calculated in this paper.}\vspace*{1cm}

The contributions from Figs. 1$(a),$ $(b),$ and $(c)$ can be obtained by
straightforward calculation. On introducing  the expansion 
\req{ai} into \eq{ham} and making use of second order 
perturbation theory and a shift to continuum integration, 
one obtains for Fig. 1$(a)$ (and its partner following  the
interchange $i\Leftrightarrow j$) the following two-body contribution to
the total energy:
\bea
\Delta E_{1a}&=&-\left(\frac{e^2}{2m_e}\right)^2\hbar\sij\ 
\dkk\frac{\cos(\vk+\vkp)\cdot\vrij}{\ok+\okp}\nn\\[.1in]
&&\quad\cdot\;\;\sll\ekek\cdot
2\ (n_{\vk\lambda}+1)(n_{\vkp\lambda\pri}+1)\ \ ,
\labeq{a}
\eea
where 
\be
n_{\vk\lambda}=\frac{1}{e^{\hbar\ok/T}-1}\ \ 
\labeq{nk}
\ee
is the Bose occupation number, and  $\vrij\equiv \ri-\rj,$

Fig. 1$(b)$ gives a  similar result:
\bea
\Delta E_{1b}&=&-\left(\frac{e^2}{2m_e}\right)^2\hbar\sij\ 
\dkk\frac{\cos(\vk+\vkp)\cdot\vrij}{\ok+\okp}\nn\\[.1in]
&&\quad\cdot\;\;\sll\ekek\cdot
\left[n_{\vk\lambda}(n_{\vkp\lambda\pri}+1)+n_{\vkp\lambda\pri}
(n_{\vk\lambda}+1)\right]\ \ ,
\labeq{b}
\eea
while Fig. 1$(c)$ gives
\bea
\Delta E_{1c}&=&\left(\frac{e^2}{2m_e}\right)^2\hbar\sij\ 
\dkk\frac{\cos(\vk-\vkp)\cdot\vrij}{\ok-\okp}\nn\\[.1in]
&&\quad\cdot\;\;\sll\ekek\cdot
(n_{\vk\lambda}-n_{\vkp\lambda\pri})\ \ .
\labeq{c}
\eea 
In order to
evaluate the thermal contribution in Eqs.~\req{a},\req{b}, and 
\req{c}, one notes that 
the integrals are effectively cut off at $k\sim 1/R,$ where $R$ is
a typical interparticle spacing, so that for massless photons 
the Bose factors are well approximated by
\be
n_{\vk\lambda}\ \simeq\  T/\hbar\ok\ \gg\  1\ \ .
\labeq{nka}
\ee
This approximation can be true even if the dispersion relation of $\ok$ 
contains an effective mass:
in that case, \eq{nka} holds as long as 
$T\gg \hbar\op.$
On substituting \req{nka} into Eqs.~\req{a},\req{b}, and \req{c}, 
I obtain at \co{T^2}
\bea
\Delta E_{1a+1b}&=&-4\left(\frac{e^2}{2m_e}\right)^2\frac{T^2}{\hbar}\sij\ 
\dkkt\frac{\cos(\vk+\vkp)\cdot \vrij}{\ok+\okp}\nn\\[.1in]
&&\qquad\cdot\ \ \ \pol\ \ ,
\labeq{abt}
\eea
 while at \co{T}   
\bea
\Delta E_{1a+1b}\pri&=&-3\left(\frac{e^2}{2m_e}\right)^2\ T\ \sij\ 
\dkkt\cos(\vk+\vkp)\cdot\vrij\nn\\[.1in]
&&\qquad\cdot\ \ \ \pol\labeq{abpt}\\[.1in]
\Delta E_{1c}&=&-\left(\frac{e^2}{2m_e}\right)^2\ T\ \sij\ 
\dkkt\cos(\vk-\vkp)\cdot\vrij\nn\\[.1in]
&&\qquad\cdot\ \ \ \pol\ \ .
\labeq{ct}
\eea
In these equations the last factor is a result of performing
the polarization sum. 

In what follows, I will ignore the $(\hat k\cdot\hat k\pri)^2$ term in the
polarization sum: it can cause no  cancellations, and greatly
complicates the algebra. With that simplifiction, the angular integrations can
easily be done, with the result
\bea
\Delta E_{1a+1b}&=&-\ \left(\frac{e^2}{4\pi^2}\right)^2\ \frac{1}{\hbar}
\ \left(\frac{T}{m_e}\right)^2\ \sij
\;\;\frac{1}{{\rij}^2}\ \nn\\[.1in]
&&\cdot\ \ \int_0^\infty\frac{k\ \sin k\rij\ dk}{\ok^2}\int_0^\infty
\frac{k\pri\ \sin {k\pri\rij}\ dk\pri}{\okp^2}\ \frac{1}{\ok+\okp}
\labeq{abr}\\[.1in]
\Delta E_{1a+1b}\pri+\Delta E_{1c}&=&-
\left(\frac{e^2}{4\pi^2}\right)^2\ 
\ \left(\frac{T}{m_e^2}\right)\ \sij
\;\;\frac{1}{{\rij}^2}\ \nn\\[.1in]
&&\cdot\ \ \int_0^\infty\frac{k\ \sin k\rij\ dk}{\ok^2}\int_0^\infty
\frac{k\pri\ \sin {k\pri\rij}\ dk\pri}{\okp^2}\ \ .
\labeq{abpcr}
\eea

For massless photons, the dispersion relation 
\be
\ok=k
\labeq{okk}
\ee
holds. In that case, the expressions \req{abr} and \req{abpcr} (with the
appropriate reinstatement of factors of $c$) become
\bea
\Delta E_{1a+1b}&=&-\left(\frac{e^2}{8\pi\hbar
c}\right)\left(\frac{T}{m_ec^2}\right)^2\ \cdot I\ \cdot\ 
\frac{1}{2}\sij \frac{e^2}{4\pi\rij}\labeq{abs}\\[.1in]
\Delta E_{1a+1b}\pri+\Delta E_{1c}&=&-\left(\frac{e^2}{8\pi}
\right)\left(\frac{T}{(m_ec^2)^2}\right)\ \cdot I\pri\ \cdot\ 
\frac{1}{2}\sij \frac{e^2}{4\pi{\rij}^2}\labeq{abpcs}\ \ ,
\eea
where 
\bea
I&=&\left(\frac{2}{\pi}\right)^2\ 
\iz\frac{\sin x\ dx}{x}\iz\frac{\sin y\ dy}{y}\frac{1}{x+y}
\simeq 0.87\nn\\[.1in]
I\pri &=&\left(\frac{2}{\pi}\right)^2\ 
\iz\frac{\sin x\ dx}{x}\iz\frac{\sin y\ dy}{y}
=1\ \ .
\labeq{ints}
\eea

It is immediately apparent that (in this massless case) 
the \co{T} contribution  (\eq{abpcs}) 
is suppressed over
\req{abs} by a factor $\hbar c/TR$, where $R$ is a typical
interparticle spacing. For the solar core, with 
\[
T\simeq 1.3 \kev,\quad R\simeq 6\times 10^9\ \mbox{cm}\ \ ,
\]
this is a factor of $10^{-18}.$ Thus, for the massless case, I focus 
on \eq{abs}, and proceed to
estimate it for the solar core environment. 

The most important aspect of \req{abs} (and also of \req{abpcs}) 
is that it
implies an attractive potential between all pairs of particles. This would
include the protons as well, but as discussed above, their contribution is
suppressed by a large mass factor $(m_e/m_{prot})^2.$ With $N$ electrons 
in a radius
$R,$ the leading thermal correction 
to the total energy is approximately given by
\be
\Delta E\ \simeq\ -\alpha^2\left(\frac{T}{m_e}\right)^2\ N^2\ 
\frac{\hbar c}{R}\ \ ,
\labeq{est}
\ee
with $\alpha$ the fine structure constant. The relevant measure of this
perturbation is its ratio  
to the thermal energy, $NT$, which is
roughly also the gravitational energy. In natural units 
$(\hbar=c=1)$, the ratio is seen to be
\be
\Delta E/(NT)\ \simeq\ -\alpha^2\left(\frac{T}{m_e}\right)\ N\ 
\frac{1}{m_eR}\ \ .
\labeq{rest}
\ee
For the solar core, $N\simeq 10^{56},$ and $R=3\times 10^{17}\ 
\mbox{KeV}^{-1},$ so that  \req{rest} is
calculated to be
\be
\left|\Delta E/(NT)\right|\simeq 10^{29}\ \ .
\labeq{nest}
\ee
This is of course totally unacceptable, and the way out is 
provided by modifying the dispersion relation for the photon in the thermal
bath.

\section{Role of the Tranverse Mass}
For $k\ll m_{pl}$ the effect of the plasma on the photons is a shift 
in the frequency dispersion relation of from \req{okk} to \req{okp}. 
When the resulting quasiparticle energy spectrum $\hbar\ok$ is introduced in
the intermediate states in Eqs.~\req{abt} or \req{abr}, the result is a Yukawa
screening of the long ranges forces.
A rough estimate of the 
modification induced in the energy by \req{okp} can be made as follows: 

As noted previously, the integrals \req{abr} and \req{abpcr} restrict 
$k\lsim R^{-1}\ll m_{pl}.$ In this case, the transverse 
photon mass in the stellar core is approximately the plasmon mass \req{op}:
\bea
m_t=\hbar\op&=&\sqrt{\frac{4\pi\alpha\ N_c}{\tfrac{4\pi}{3}R_c^3m_e}}
\nn\\[.1in]
&\simeq& 400\ \mbox{eV}\ \ ,
\labeq{mpl}
\eea
where $R_c$ is the core radius and $N_c$ is the number of electrons in the
core.
Since $T/\hbar\op\simeq 3.3,$ one may justifiably use
the approximation \req{nka} and the simplified form \req{abr} 
for order-of-magnitude discussion. The integral in \req{abr} 
may be estimated in the following manner: the
introduction of the dispersion relation \req{okp} gives
contributions of
the form \req{est}, but limited to subvolumes of radius $R_p\le m_t^{-1}, $
with the
number of particles $N_p$ in each of these subvolumes given by 
\be
N_p=N_c\ (R_p/R_c)^3\ \ .
\labeq{enp}
\ee
The number of such volumes in the core is $N_c/N_p,$ so that the total
contribution to the \co{T^2}energy is (from \req{est} and \req{enp})
\bea
\Delta E_{1a+1b}\ (\mbox{\em screened})&=&
 -\left(\frac{N_c}{N_p}\right)\ \alpha^2\ \left(\frac{T}{m_e}\right)^2\ N_p^2\ 
\frac{1}{R_p}\labeq{estsca}\\[.1in]
&=&\left(\frac{R_p}{R_c}\right)^2\ \cdot\  
\Delta E_{1a+1b}\ (\mbox{\em unscreened})\ \ ,
\labeq{estscb}
\eea
The ratio of interaction energy/thermal energy then becomes (omitting $-$
signs)
\bea
\Delta E_{1a+1b}\ (\mbox{\em screened})/(N_cT)&=&(R_p/R_c)^2\times 10^{29}
\nn\\[.1in]
&=&7\times 10^{-35}\times 10^{29}\nn\\
&\simeq&7\times 10^{-6}\ \ ,
\labeq{estscnum}
\eea
which renders this effect negligible. A bit more insight is afforded by
an algebraic approach: substitute into \eq{estsca} the
relation \req{enp} and $R_p=m_t^{-1}$ with $m_t$ as given in 
\eq{mpl}. The result is 
\be
\Delta E_{1a+1b}\ (\mbox{\em screened})/(N_cT)=\frac{\alpha}{3}\ \cdot
\frac{T}{m_e}\ \ .
\labeq{estscalg}
\ee
This shows that the suppression depends only on the non-relativistic property
of the electrons, which was an assumption of the calculation in the first
place. 

The \co{T} contribution \req{abpcr} may be similarly estimated: although it
is greatly suppressed over the \co{T^2} term for massless photons, this is not
so for quasiparticles. Following the same steps which led to \eq{estscalg}, I
find
\be
\left(\Delta E_{1a+1b}\pri+\Delta E_{1c}\right)(\mbox{\em screened})/(N_cT)
\simeq 
\frac{\alpha}{3}\
\cdot \frac{m_t}{m_e}\ \ ,
\labeq{esttalg}
\ee
which is again small (of \co{10^{-6}}), but not greatly suppressed over the
\co{T^2} contribution.

\section{Summary and Concluding Remarks}

\ni{1}In this paper I have calculated corrections to the static two-body
Coulomb potential in a thermal photon bath. For
massless transverse photons, the interaction energy per particle is 
many orders of magnitude larger than $kT.$ The screening provided by 
introducing  a photon energy
spectrum appropriate to  quasiparticles  with mass
$m_{pl}=\hbar\sqrt{e^2n_e/m_e}$ is sufficient to suppress this effect 
in the solar
core to negligible significance. In this sense, the argument can be made that
quantized transverse quasiparticles are a necessary component for a successful
description of physics of the stellar core. 

\ni{2} {\em The Kinetic Terms.} It is not difficult to see that 
for massless photons and  to
leading order in temperature $T,$ some of the individual
kinetic terms (such as Figs. 1$(d)$ and $1(f)$) are infrared divergent. 
Such divergences will presumably cancel when
all contributions are taken into account, in a manner similar to that 
found by
Tryon \cite{Tryon} in his discussion of scattering from an external 
potential. However, in Coulomb gauge, even the finite residual part will be
distinguished from the static part calculated in this work by having 
explicit
momentum dependence, and hence such contributions can be considered
independently of those in this paper.
Thus, the results of the calculations in this paper should be  sufficient for
assessing the necessity of the modified dispersion relation \req{okp}.

\ni{3}Because the thermal
corrections calculated here, being of \co{e_i^2e_j^2}, are attractive 
between {\em all} pairs \{$ij$\} of particles, the usual 
Debye screening plays no role in alleviating the energy problem discussed.
It is the transverse plasmon mass which is the relevant mitigating factor.

\subsection*{Acknowledgement}This research was supported in part by Grant No.
PHY-9722044 from the National Science Foundation.
%
%

\begin{thebibliography}{99}
\bb{Tonks}L.~Tonks and I.~Langmuir, {\em Phys.~Rev.}~{\bf 33} (1929) 195.
\bb{Powell}C.~J.~Powell and J.~B.~Swan, {\em Phys.~Rev.}~{\bf 115} (1959) 869;
{\bf 116} (1959) 81.
\bb{Tsytovich}V.~N.~Tsytovich, {\em Sov.~Phys.~JETP} {\bf 13} (1961) 1249.
\bb{Ruderman}J.~B.~Adams, M.~A.~Ruderman, and C.-H.~Woo, 
{\em Phys.~Rev.} {\bf 129} (1963) 1383. 
\bb{Braaten}E.~Braaten and D.~Segel, \prd{48} (1993) 1478. This paper presents a
calculation of $\omega_t(k)$ over a wide range of wave number $k.$
\bb{Dolan}L.~Dolan and R.~Jackiw, \prd{9} (1974) 3320; 
S.~Weinberg, \prd{9} (1974) 3357; B.~Freedman and L.~McLerran,
\prd{16} (1977) 1130, 1169; J.~Kapusta, \np{148} (1979) 461; 
D.~J.~Gross, R.~D.~Pisarski, and L.~G.~Yaffe, {\em
Rev.~Mod.~Phys.} {\bf 53} (1981) 43.
\bb{Tryon}E.~P.~Tryon, \prl{32} (1974) 1139.
\eb\ed